\documentclass[journal]{IEEEtran}
\IEEEoverridecommandlockouts
\usepackage{booktabs}
\usepackage{threeparttable}
\usepackage{amsmath,graphicx,cite,amssymb}
\usepackage{amsfonts,multirow,bm,array,setspace}
\usepackage{graphicx,float,cite,amssymb}
\usepackage{multirow,bm,array,setspace}
\usepackage{textcomp}
\usepackage{psfrag}
\usepackage{color}
\usepackage{pstricks}
\usepackage{bbm}
\usepackage{algorithm,algpseudocode}
\usepackage{cuted}
\usepackage{subfigure}
\usepackage{xpatch}

\def\BibTeX{{\rm B\kern-.05em{\sc i\kern-.025em b}\kern-.08em
    T\kern-.1667em\lower.7ex\hbox{E}\kern-.125emX}}

\makeatletter
\newcommand{\distas}[1]{\mathbin{\overset{#1}{\kern\z@\sim}}}%
\newsavebox{\mybox}\newsavebox{\mysim}
\newcommand{\distras}[1]{%
	\savebox{\mybox}{\hbox{\kern1pt$\scriptstyle#1$\kern1pt}}%
	\savebox{\mysim}{\hbox{$\sim$}}%
	\mathbin{\overset{#1}{\kern\z@\resizebox{\wd\mybox}{\ht\mysim}{$\sim$}}}%
}

\ExplSyntaxOn
\cs_new:Npn \bibColoredItems #1#2
  {
    \clist_map_inline:nn {#2} { \cs_new:cpn {bib@colored@##1} {#1} } 
  }
\ExplSyntaxOff

\newcommand\bib@setcolor[1]{%
  \ifcsname bib@colored@#1\endcsname
    \expandafter\color\expandafter{\csname bib@colored@#1\endcsname}
  \else
    \normalcolor
  \fi
}

\xpatchcmd\@bibitem
  {\item}
  {\bib@setcolor{#1}\item}
  {}{\PatchFailed}

\xpatchcmd\@lbibitem
  {\item}
  {\bib@setcolor{#2}\item}
  {}{\PatchFailed}

\begin{document}

\title{Power Allocation for Finite-Blocklength IR-HARQ}

\author{
        \IEEEauthorblockN{
        Wenyu Wang, \IEEEmembership{Graduate Student Member,~IEEE},  Minhao Zhu, \IEEEmembership{Graduate Student Member,~IEEE},\\ Kaiming Shen, \IEEEmembership{Senior Member,~IEEE},
        Zhaorui Wang, \IEEEmembership{Member,~IEEE}, and Shuguang Cui, \IEEEmembership{Fellow,~IEEE}} 
        
        \thanks{
            Manuscript submitted on \today. The work was supported in part by the Basic Research Project No. HZQB-KCZYZ-2021067 of Hetao Shenzhen-HK S\&T Cooperation Zone and in part by the NSFC under Grant 92167202. \emph{(Corresponding authors: Kaiming Shen and Zhaorui Wang.)}
            
            The authors are with the School of Science and Engineering, the Future Network of Intelligence Institute, The Chinese University of Hong Kong, Shenzhen, China (e-mail: \{wenyuwang, minhaozhu\}@link.cuhk.edu.cn; \{shenkaiming, wangzhaorui, shuguangcui\}@cuhk.edu.cn). 
            }
}

\maketitle

\begin{abstract}
This letter concerns the power allocation across the multiple transmission rounds under the Incremental Redundancy Hybrid Automatic Repeat reQuest (IR-HARQ) policy, in pursuit of an energy-efficient way of fulfilling the outage probability target in the finite-blocklength regime. We start by showing that the optimization objective and the constraints of the above power allocation problem all depend upon the outage probability. The main challenge then lies in the fact that the outage probability cannot be written analytically in terms of the power variables. To sidestep this difficulty, we propose a novel upper bound on the outage probability in the finite-blocklength regime, which is much tighter than the existing ones from the literature. Most importantly, by using this upper bound to approximate the outage probability, we can recast the original intractable power allocation problem into a geometric programming (GP) form---which can be efficiently solved by the standard method.
\end{abstract}

\section{Introduction}
This letter seeks the optimal power allocation for the Incremental Redundancy Hybrid Automatic Repeat reQuest (IR-HARQ) scheme \cite{ldpcirharq,caire} in the finite-blocklength regime. Upon error detection, differing from the classic HARQ that directly resends the entire packet, the IR-HARQ would transmit the coded redundancy bits to achieve the \emph{mutual information accumulation} \cite{caire} and thereby facilitate the message decoding. To account for the limited battery life of those small devices of the Internet of Things (IoT) and the industrial automation, this paper aims at an energy-efficient power control for the IR-HARQ in the finite-blocklength regime. 

A major challenge in designing the power control scheme lies in characterizing the outage probability for each retransmission of IR-HARQ. In the realm of the conventional IR-HARQ wherein the blocklength is assumed to be infinitely large, the early attempt \cite{lanemanbound} suggests bounding the complicated outage probability function from above in closed-form, followed by the recent works \cite{shen,wenyubound} that aim to improve the classic upper bound in \cite{lanemanbound}. However, the packets adopted in the IoT applications are often fairly short for the real-time transmission purpose. Consequently, the outage probability characterization assuming infinitely blocklength as in the classic information theory can incur large distortion and thus cause huge performance loss \cite{urllcfinite}. 


The early works \cite{buckingham,polyanskiy} about the finite blocklength focuses on the one-shot transmission over the additive white Gaussian noise (AWGN) channel, aiming to either compute the outage probability given the number of bits or compute the channel capacity given the outage probability. While some existing works such as \cite{predictor} extend the above results to other types of channel, many other works \cite{throughput1,throughput2,energy1,he} propose the extensions to the IR-HARQ. In principle, by following how the authors of \cite{buckingham,polyanskiy} approximated the key variable $W$ (which will be specified in Section II of this paper), \cite{throughput1,throughput2,energy1,he} obtain the outage probability evaluation for the IR-HARQ given the fixed transmit powers. All of \cite{throughput1,throughput2,energy1,he} adopt a quasi-static channel model, i.e., the channels are randomly generated at the beginning of each sub-block (i.e., each round of transmission of the IR-HARQ) and are fixed within the same sub-block. Differing from \cite{throughput1,throughput2,energy1,he} that assume equal sub-block length, a more recent work \cite{avranas} allows the sub-block length to vary and considers a joint optimization of sub-block length and transmit power. It is worth pointing out that all the above works assume that the (statistical) channel state information (CSI) is precisely known at the transmitter. In contrast, another recent work \cite{oCSIfading} examines the outage probability of the IR-HARQ in the presence of the CSI error.

The present work can be distinguished from these existing efforts in two respects. First, their problem formulations are different. Notice that \cite{throughput1,throughput2,energy1} only characterize the outage probability of the IR-HARQ when the transmit power is fixed. To the best of our knowledge, very few existing works consider power allocation for the IR-HARQ in the finite-blocklength regime, even though the power allocation for the IR-HARQ has been considered extensively \cite{diffpow_lambound1,diffpow_lambound2} assuming an infinitely long blocklength. 
In particular, \cite{avranas} considers optimizing power allocation in conjunction with blocklength, and yet assumes fixed channel throughout, whereas our work adopts the quasi-static fading channel model in \cite{throughput1,throughput2,energy1}. Second, the ways of evaluating the outage probability are different. The previous works \cite{predictor,throughput1,throughput2,energy1,he,avranas} are typically based on the outage probability approximation method in \cite{buckingham}. In contrast, this work proposes a new upper bound to the actual outage probability of the finite-blocklength IR-HARQ. This new upper bound brings two benefits: (i) it yields a tighter approximation of the actual outage probability in the finite-blocklength regime; (ii) it converts the original problem to a geometric programming (GP) form.

\section{System Model}
\begin{figure}
    \centering
    \centerline{\includegraphics[width=1.0\linewidth]{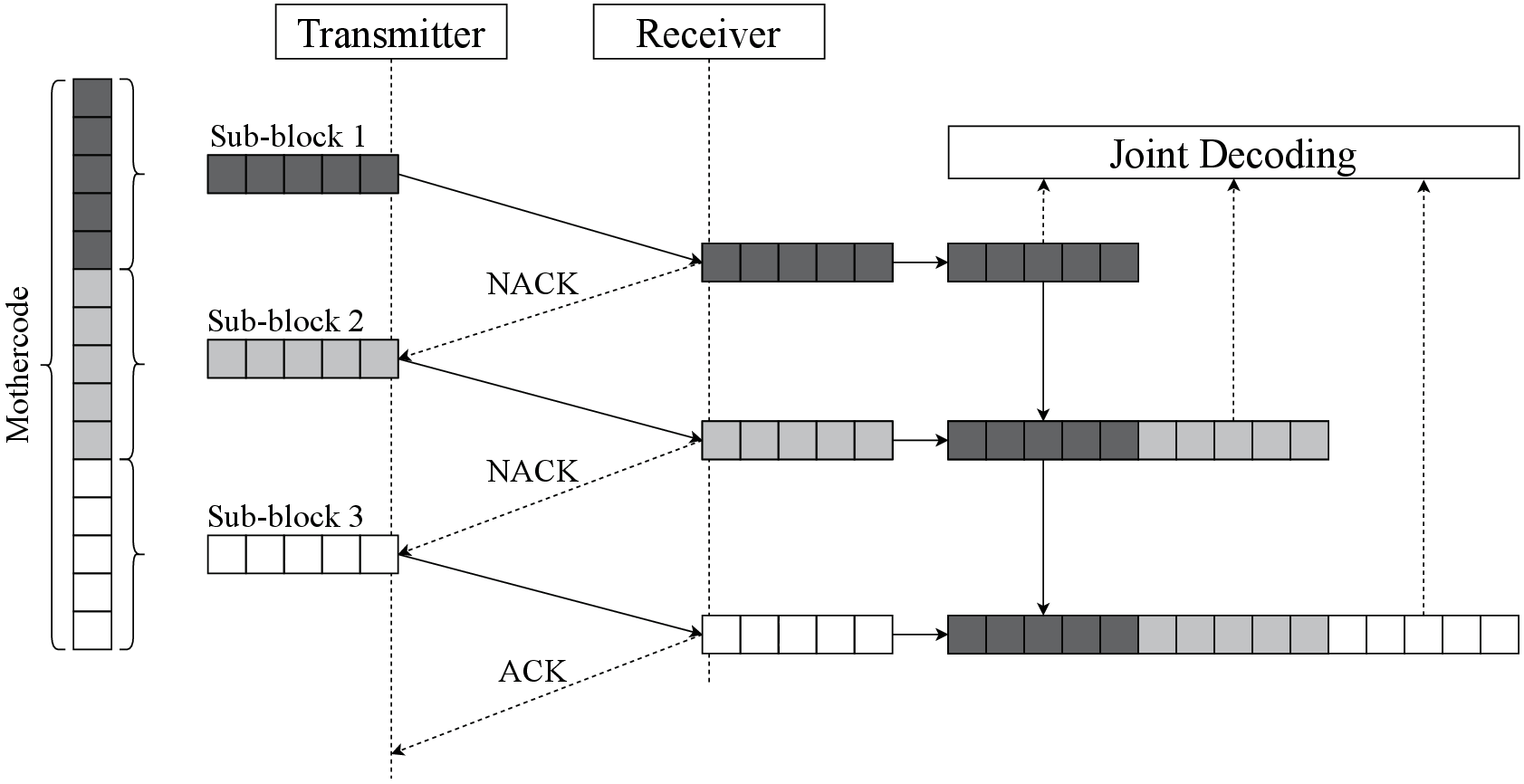}}
    \caption{The procedure of the IR-HARQ scheme.}
    \label{fig:IR_HARQ}
\end{figure}
Consider delivering a message of $q$ nats\footnote{We measure information in nats and use log base $e$ throughout the paper.} by the IR-HARQ scheme. Assume that the message is encoded into a sequence of $L$ symbols $\boldsymbol{X} = (X_1,X_2,\dots,X_L)$, where each $X_l\in\mathbb C$. The $L$ symbols (or channel uses) are further divided into $M$ sub-blocks, each corresponding to a transmission round of the IR-HARQ. We assume the variable sub-block lengths and denote by $L_m$ the number of channel uses assigned to sub-block $m$. It is required that $\sum^M_{m=1}L_m=L$. Moreover, the total blocklength across the first $m$ sub-blocks is written as
\begin{equation}
    \widetilde{L}_m = \sum^m_{j=1} L_j.
\end{equation}

We adopt a quasi-static fading channel model \cite{diffpow_lambound1,diffpow_lambound2,throughput1,throughput2,energy1} between the transmitter and the receiver, i.e., the channel is randomly generated at the beginning of each sub-block and is fixed within the sub-block. Let $h_m\in\mathbb C$ be the channel of sub-block $m=1,2,\ldots,M$. More specifically, each $h_m$ is modeled as a Rayleigh fading:
\begin{equation}
    h_m = \sqrt{\beta}g_m,
\end{equation}
where the positive factor $\beta>0$ is the large-scale fading component and is fixed throughout the $M$ sub-blocks, and the standard complex Gaussian random variable $g_m\sim\mathcal{CN}(0, 1)$ is the small-scale fading i.i.d. distributed across the sub-blocks.

The procedure of the IR-HARQ scheme is illustrated in Fig.~\ref{fig:IR_HARQ}. At the beginning, the packet in sub-block 1 is transmitted. If the receiver can successfully decode the $q$-nat message (e.g., according to the parity check bits), then it sends feedback \textsf{ACK} to the transmitter so that the transmission finishes. Otherwise, the receiver sends feedback \textsf{NACK} to notify the transmitter of the decoding error. Upon receiving \textsf{NACK}, the transmitter enters the second round of transmission of the IR-HARQ and sends the packet in sub-block 2. At the receiver side, the previously received packet and the current packet are decoded jointly so as to achieve the \emph{incremental redundancy} effect. Again, the \textsf{ACK}/\textsf{NACK} feedback is performed after the joint decoding. Based on the feedback, the transmitter decides to either finish the transmission or enter the next round of transmission. In particular, after $M$ rounds of transmission, the transmission must terminate regardless of the \textsf{ACK}/\textsf{NACK} feedback. If the receiver still fails to recover the $q$-nat message after $M$ sub-blocks, then it declares an \emph{outage}.

We now further evaluate the outage probability. By convention, assume that the transmit symbols within sub-block $m$ are drawn i.i.d. according to a complex Gaussian distribution $\mathcal{CN}(0, p_m)$, where $p_m$ is the transmit power allocated to sub-block $m$. The $l$th received symbol $Y_l$, assuming that it is contained in sub-block $m$, can be computed as
\begin{align}
    Y_l=h_mX_l+N_l,
\end{align}
where the additive noise $N_l$ has a complex Gaussian distribution $\mathcal{CN}(0,N_0)$, where $N_0$ is the noise power. Let $(x_l,y_l)$ be a realization of $(X_l,Y_l)$ for the $l$th channel use. We write
\begin{align}
\boldsymbol{x}^{\widetilde{L}_m}&=(x_1,x_2,\dots,x_{\widetilde{L}_m}),
\end{align}
which is the sequence of all the transmit symbols till the end of sub-block $m$. Similarly, we write
\begin{align}
    \boldsymbol{y}^{\widetilde{L}_m}&=(y_1,y_2,\dots,y_{\widetilde{L}_m}).
\end{align}
To further obtain an achievable rate across the first $M$ sub-blocks, we can evaluate the \emph{mutual information density} between $\boldsymbol{x}^{\widetilde{L}_m}$ and $\boldsymbol{y}^{\widetilde{L}_m}$ according to \cite{buckingham,laneman2006distribution} as
\begin{align}   
i(\boldsymbol{x}^{\widetilde{L}_m};\boldsymbol{y}^{\widetilde{L}_m})&=\frac{1}{\widetilde{L}_m}\sum^{\widetilde{L}_m}_{l=1}i(x_l;y_l)=\sum^m_{j=1}\frac{L_j}{\widetilde{L}_m}(C_j+W_j),
\label{i x^nL y^nL}
\end{align}
where 
\begin{equation}
\label{C_j}
    C_j = \log\left(1+\frac{|h_j|^2p_j}{N_0}\right)
\end{equation}
and
\begin{align}
\label{W_j}
    W_j = \frac{1}{L_j}\sum^{\widetilde{L}_j}_{l=\widetilde{L}_{j-1}+1}\left(\frac{|Y_l|^2}{\beta|g_j|^2p_j+N_0}-\frac{|Z_l|^2}{N_0}\right).
\end{align}
In particular, as $L_j\rightarrow\infty$, $W_j$ tends to zero and consequently the achievable rate in \eqref{i x^nL y^nL} reduces to Shannon's formula.
As shown in \cite{laneman2006distribution,buckingham}, $W_j\in\mathbb R$ admits a Bessel-$K$ distribution with the probability density function (PDF)
\begin{multline}
f_{W_j}(w)=\frac{2^{1-L_j}}{\sqrt{\pi}\sigma_{W_j}(L_j-1)!}\times\left(\frac{\sqrt{2}|w|}{\sigma_{W_j}}\right)^{L_j-\frac{1}{2}}\\\times\mathcal K_{L_j-\frac12}\left(\frac{\sqrt{2}|w|}{\sigma_{W_j}}\right),\label{condpdfW}
\end{multline}
where 
\begin{align}
    \sigma_{W_j}=\sqrt{\frac{2|h_j|^2p_j}{|h_j|^2p_j+N_0}}
\end{align}
and $\mathcal K_{L_j-\frac{1}{2}}(\cdot)$ is the modified Bessel function of the second kind:
\begin{align}
    \mathcal K_{L_j-\frac{1}{2}}(t)=\sqrt{\frac{\pi}{2t}}e^{-t}\sum^{L_j-1}_{l=0}\frac{(L_j+l-1)!\,(2t)^{-l}}{(L_j-l-1)!\,l!}.
\end{align}
For the special case of $L_j=1$, the above $\sigma_{W_j}$ boils down to the standard variance of ${W_j}$ when $L_j=1$.
As in \cite{buckingham,throughput2}, assume that the receiver fails to decode the message after $m$ sub-blocks if
\begin{equation}
\label{outage:HARQ}
    i(\bm x^{\widetilde{L}_m}; \bm y^{\widetilde{L}_m}) < R_m,
\end{equation}
where $R_m= q/{\widetilde{L}_m}$. Letting
\begin{equation}
    U_j = L_j(C_j + W_j),
    \label{Um}
\end{equation}
we can rewrite \eqref{outage:HARQ} as $\sum^m_{j=1} U_j < q$. In other words, the joint decoding of the first $m$ sub-blocks fails if
\begin{equation}
\label{En}
    \mathcal{E}_m = \left\{ \sum^m_{j=1} U_j < q\right\}.
\end{equation}
Furthermore, an outage is said to occur at the end of sub-block $m$ if the joint decoding has failed after each sub-block thus far \cite{ldpcirharq,caire}, so the outage probability after $m$ sub-blocks is defined to be
\begin{align}
\label{Qn_En}
    \mathrm{Pr}_{\mathrm{out}}(m)=\Pr\left[\mathcal{E}_{1}\cap\mathcal{E}_{2}\cap\ldots\cap\mathcal{E}_{m}\right].
\end{align}
Recall that the whole transmission encounters an error if an outage persists after all the $M$ sub-blocks have been used, so the ultimate outage probability equals $\mathrm{Pr}_{\mathrm{out}}(M)$.

We now evaluate the expected power consumption with $\mathrm{Pr}_{\mathrm{out}}(m)$. Notice that sub-block 1 is always used, and also that sub-block $m$ is used if and only if the previous sub-block $m-1$ encounters an outage, so the expected energy consumption can be computed as
\begin{align}
    E &= p_1L_1 + p_2L_2\mathrm{Pr}_{\mathrm{out}}(1) + \ldots + p_ML_M\mathrm{Pr}_{\mathrm{out}}(M-1) \label{eq:E}
\end{align}
Let $\epsilon>0$ be the ultimate outage probability constraint and let $P>0$ be the power consumption constraint. We seek the optimal power allocation $(p_1,\ldots,p_M)$ that minimizes the expected energy consumption under the above constraints:
\begin{subequations}
\label{prob}
\begin{align}
    \underset{(p_1,\dots,p_M)}{\text{minimize}}&\quad E\\
    \text{subject to}& \quad \mathrm{Pr}_{\mathrm{out}}(M)\leq\epsilon\\
    &\quad 0\leq p_m\leq P.
\end{align}
\end{subequations}
In the above problem, the transmitter already knows the channel statistics beforehand, and aims to pre-allocate the power allocation before the IR-HARQ takes place. We remark that similar power allocation problems have been considered in \cite{diffpow_lambound1,diffpow_lambound2} and \cite{avranas}. Nevertheless, \cite{diffpow_lambound1,diffpow_lambound2} assumes an infinitely long blocklength, while \cite{avranas} assumes a fixed channel across sub-blocks.

\section{Power Allocation by Bounding and GP}\label{sec:pro}

\subsection{Existing Approximation of Outage Probability}
Problem \eqref{prob} is difficult to tackle directly because the outage probability $\mathrm{Pr}_{\mathrm{out}}$ in \eqref{Qn_En} cannot be written analytically. Write the \emph{pathloss-to-noise ratio} as
\begin{equation}
    S=\beta/N_0. \label{PNR}
\end{equation}
In the existing literature, most works \cite{throughput1,throughput2,energy1} rely on \cite{polyanskiy} to approximate ${\mathrm{Pr}}_{\mathrm{out}}(m)$ as,
\begin{equation}\label{eq:Qnapprox}
    \mathrm{Pr}'_{\mathrm{out}}(m) = \mathbb{E}_{v_m}\left[\frac{1}{\sqrt{2\pi}}\int^{\infty}_{v_m}e^{-\frac{t^2}{2}}dt\right],
\end{equation}
where the random variable $v_m$ depends on $S$, $(p_1,\ldots,p_m)$, and $(g_1,\ldots,g_m)$ as
\begin{equation}\label{eq:vm}
    v_m = \frac{\sum^m_{j=1}L_j\log(1+Sg_jp_j)-q}{\sqrt{\sum^m_{j=1}\frac{L_j(2+Sg_jp_j)Sg_jp_j}{(Sg_jp_j+1)^2}}}.
\end{equation}
The above approximate outage probability $\mathrm{Pr}'_{\mathrm{out}}(m)$ is sufficient for characterizing the finite-blocklength performance given fixed powers \cite{throughput1,throughput2,energy1}, but using it to replace $\mathrm{Pr}_{\mathrm{out}}(m)$ in \eqref{prob} cannot render the power allocation problem easier. Consequently, the previous work \cite{he} has to further relax the above approximation  dramatically in order to perform GP optimization. We therefore seek a better approximation.

\subsection{Proposed Approximation of Outage Probability}

Before proceeding to the new approximation, we first take a closer look at the actual outage probability $\mathrm{Pr}_{\mathrm{out}}(m)$.
We start by specifying the computation in \eqref{En} as
\begin{multline}
\mathrm{Pr}_{\mathrm{out}}(m) = \int^{q}_{-\infty}\int^{q-u_1}_{-\infty}\cdots\int^{q-\sum^{m-1}_{j=1}u_j}_{-\infty}\prod^m_{j=1}f_{U_j}(u_j)\\
du_m d{u_{m-1}}\cdots d{u_1},
    \label{Qn}
\end{multline}
where $f_{U_j}(u_j)$ is the PDF of $U_j$. Next, we aim to compute $f_{U_j}(u_j)$. First, for the Rayleigh fading channel $h_m$, the resulting $C_m$ in \eqref{C_j} has the PDF
\begin{align}
\label{f_Cn}
    f_{C_m}(c) = \frac{e^c}{S p_m}\cdot\exp\bigg(-\frac{(e^c-1)}{S p_m}\bigg).
\end{align}
Moreover, the PDF of $W_m$ conditioned on $C_m$ in \eqref{W_j} is
\begin{multline}
f_{W_m|C_m = c}(w) = \frac{2^{L_m+\frac{1}{2}}e^{\frac{c}{2}}(e^c-1)^{-\frac{1}{2}}}{\sqrt{\pi}(L_m-1)!}\times\bigg(\frac{e^{\frac{c}{2}}|w|}{\sqrt{e^c-1}}\bigg)^{L_m-\frac{1}{2}}\\
\times \mathcal{K}_{L_m-\frac{1}{2}}\bigg(\frac{e^{\frac{c}{2}}|w|}{\sqrt{e^c-1}}\bigg).
\end{multline}
We then obtain the joint PDF of $(W_m,C_m)$ as
\begin{equation}
\label{f_WC}
    f_{W_m,C_m}(w,c) = f_{W_m|c}(w) f_{C_m}(c).
\end{equation}
As a result, the PDF of $U_m$ can now be computed as
\begin{align}
\label{f_U}
    f_{U_m}(u) = \int_{0}^{+\infty}f_{W_m,C_m}\left(\frac{u}{L_m}-c,c \right)dc.
\end{align}
It can be seen that $f_{C_m}(c)$ in \eqref{f_Cn} is impacted by $p_m$ alone but then $\mathrm{Pr}_{\mathrm{out}}(m)$ in \eqref{Qn} is jointly impacted by $(p_1,p_2,\ldots,p_m)$ through $f_{W_m,C_m}(w,c)$. The connection between $(p_1,p_2,\ldots,p_m)$ and $\mathrm{Pr}_{\mathrm{out}}(m)$ is quite complicated, causing \eqref{prob} difficult to solve directly.

This work suggests bounding $\mathrm{Pr}_{\mathrm{out}}(m)$ from above and then using the upper bound as an approximation. We start by bounding ${f}_{C_m}(c)$ in \eqref{f_Cn} from above as
\begin{equation}
  f_{C_m}(c) \le  \hat{f}_{C_m}(c) = \frac{e^c}{S p_m}\cdot\exp\bigg(-\frac{(e^c-1)}{S P}\bigg).
\end{equation}
It turns out that the above upper bound carries over to the PDF of $U_m$ by replacing $f_{C_m}(c)$ with $\hat{f}_{C_m}(c)$ in \eqref{f_WC}. The PDF of $U_m$ is then bounded as
\begin{align}
  f_{U_m}(u) \le  \hat{f}_{U_m}(u)=\frac{\phi_m(u)}{p_m},
\end{align}
where
\begin{multline}\label{eq:phim}
    \phi_m(u)=\int_{0}^{\infty}f_{W_m|c}\left(\frac{u}{L_m}-c\right)\\
    \cdot\frac{2^c\ln{2}}{S}\cdot\exp\bigg(-\frac{2^c-1}{S P}\bigg)dc.
\end{multline}
Notably, even though $\hat{f}_{U_m}(u)$ still looks complicated due to $\phi_m(u)$, it is a simple function of $p_m$. 

\begin{figure}
    \centering
    \centerline{\includegraphics[width=0.9\linewidth]{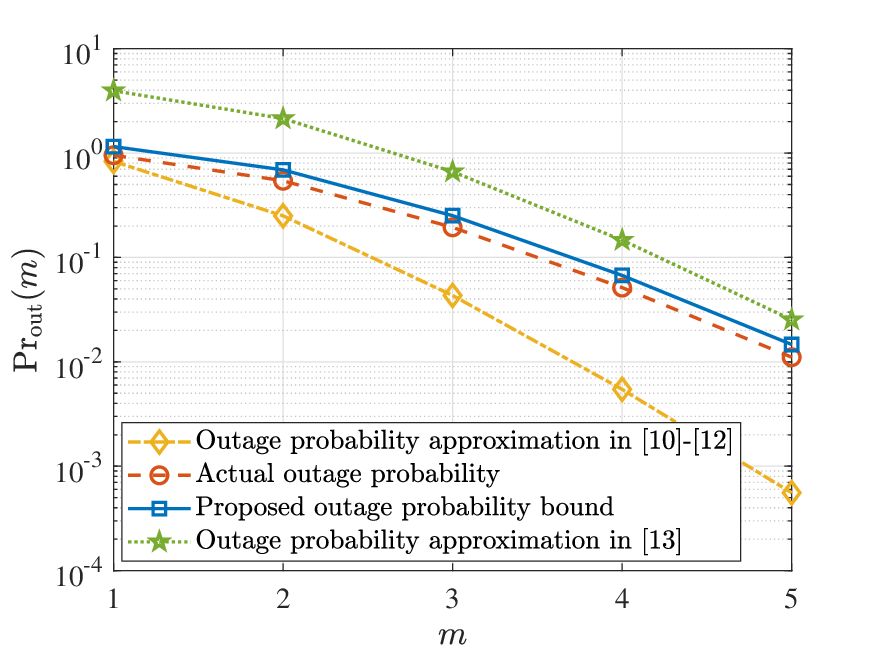}}
    \caption{Approximations of $\mathrm{Pr}_{\mathrm{out}}(m)$ when $M=5$, $q=40\log(2)\text{ nats}$, $L=50$, $S=5$, and each $p_m=0.8$.}
    \label{fig:Qnvsn}
\end{figure}

\begin{figure}
    \centering
    \centerline{\includegraphics[width=0.9\linewidth]{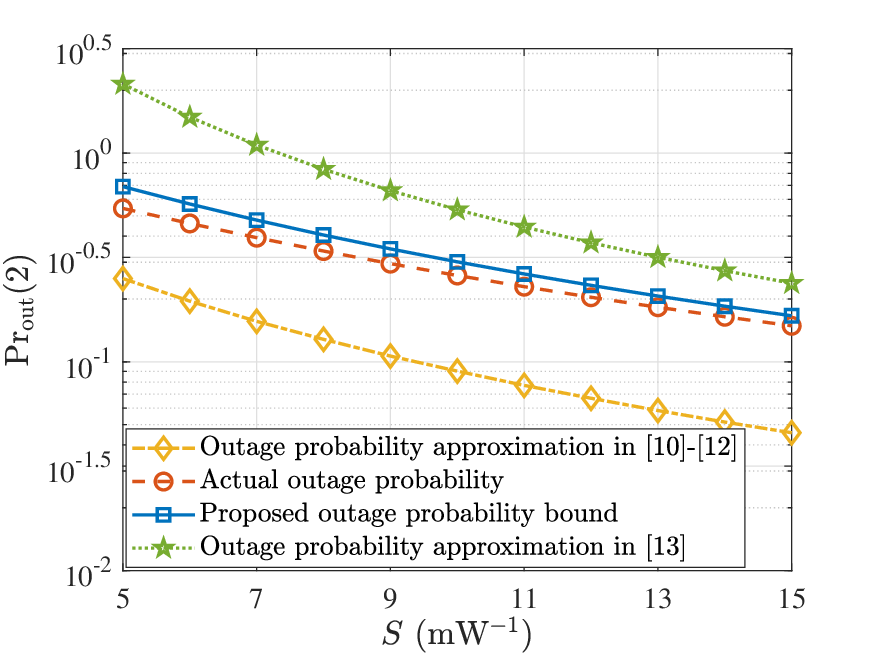}}
    \caption{Approximation of $\mathrm{Pr}_{\mathrm{out}}(2)$ vs. $S$ when $M=5$, $q=40\log(2)\text{ nats}$, $L=50$, and each $p_m=0.8$.}
    \label{fig:Q3vsbeta}
\end{figure}

Substituting ${f}_{C_m}(c)$ with $\hat{f}_{C_m}(c)$ in \eqref{Qn} further yields an upper bound on the outage probability $\mathrm{Pr}_{\mathrm{out}}(m)$:
\begin{align}
\label{proposed_upper_bound}
 \mathrm{Pr}_{\mathrm{out}}(m) \le   \widehat{\mathrm{Pr}}_{\mathrm{out}}(m)&=
    \frac{1}{\prod^n_{m=1}p_m}A_m,
\end{align}
where
\begin{multline}\label{eq:Am}
A_m =     \int^{q}_{-\infty}\int^{q-u_1}_{-\infty}\dots\int^{q-\sum^{m-1}_{j=1}u_j}_{-\infty}\prod^m_{j=1}\phi_j(u_j)\\
du_mdu_{m-1}\dots du_1.
\end{multline}
Again, although $\widehat{\mathrm{Pr}}_{\mathrm{out}}(m)$ in \eqref{proposed_upper_bound} appears complicated, it is a simple function of $(p_1,\ldots,p_m)$. Fig.~\ref{fig:Qnvsn} and Fig.~\ref{fig:Q3vsbeta} show that the new approximation is much tighter than the existing approximations,
with the actual outage probability computed as in \eqref{Qn}.

\subsection{GP-Based Power Allocation Scheme}

With each $ \mathrm{Pr}_{\mathrm{out}}(m)$ replaced by $\widehat{\mathrm{Pr}}_{\mathrm{out}}(m)$, the power allocation problem \eqref{prob} is converted to
\begin{subequations}
\label{prob:new}
\begin{align}
    \underset{(p_1,\dots,p_M)}{\text{minimize}}&\quad p_1L_1+\sum_{m=2}^{M}\frac{A_{m-1}L_m p_m}{\prod_{j=1}^{m-1} p_j}\\
    \text{subject to}& \quad \frac{A_M}{\prod^M_{m=1}p_m}\leq\epsilon\\
    &\quad 0\leq p_m\leq P.
\end{align}
\end{subequations}
Furthermore, notice that the new problem \eqref{prob:new} is a GP problem, so it can be efficiently solved by the standard method. Most importantly, since in the new problem we only approximate the actual outage probability as its upper bound, the solution of \eqref{prob:new} is guaranteed to be feasible in the original problem \eqref{prob}, i.e., it must satisfy the ultimate outage probability constraint and the power constraints. 

The overall complexity of the proposed GP algorithm is $\mathcal{O}(M^2(\varphi^{-2}+\varphi^{-1}\vartheta^{-1})+M^{3.5}\log(\varepsilon^{-1}))$, where $\varphi$ is the required precision to compute the integration in \eqref{eq:phim}, $\vartheta$ is the required precision to compute the integration in \eqref{eq:Am}, and $\varepsilon$ is the required precision to solve the GP problem.

\section{Numerical Results}
This section numerically validates the performance of the proposed GP-based power allocation scheme. The simulation setting follows: the power constraint $P=1$ mW, the number of sub-blocks $M=5$, the outage probability constraint $\epsilon=10^{-4}$ as in \cite{DAHLMAN2018251}, the message size $q=40\log(2)$ nats, and the total number of channel uses $L=50$. Following \cite{throughput1,throughput2,energy1}, we assume equal sub-blocklength and thus each $L_m=L/M=10$. Other parameters are chosen differently and are specified later on. As discussed earlier, to the best of our knowledge, only \cite{avranas} in the existing literature concerns the power allocation for the IR-HARQ in the finite-blocklength regime, but \cite{avranas} assumes fixed channel across the sub-blocks and hence does work account for our problem case. 
For the comparison purpose, we consider the following two benchmarks: 
\begin{itemize}
    \item \emph{Max Power Scheme:} Assume that each sub-block uses the max power $P$, i.e., $p_m=P$ for each $m$.
    \item \emph{Infinite Blocklength Method:} Assume infinitely long blocklength and use the existing method in \cite{shen}.
\end{itemize}

\begin{figure}
    \centering    \centerline{\includegraphics[width=0.9\linewidth]{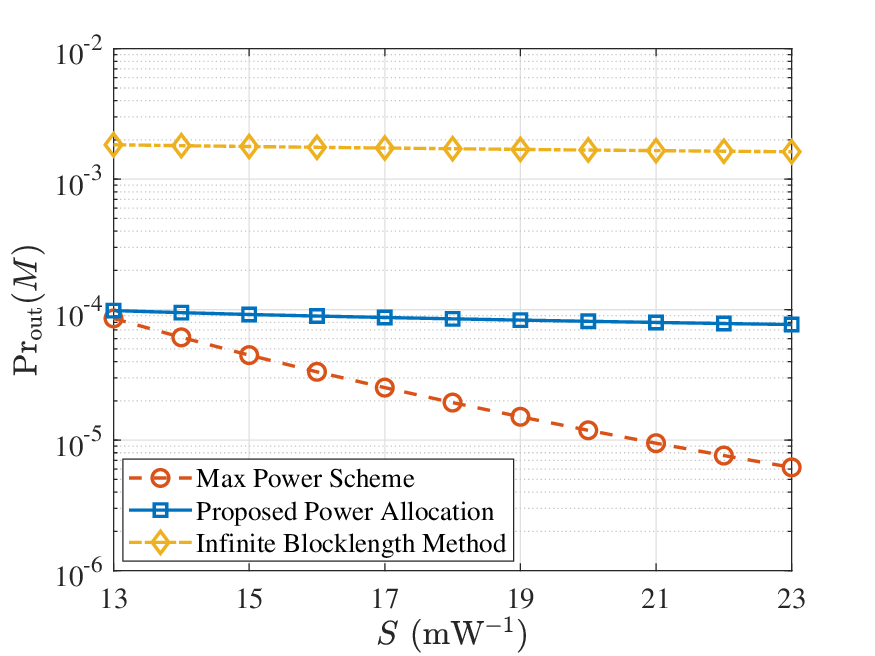}}
    \caption{Ultimate outage probability $\mathrm{Pr}_{\mathrm{out}}(M)$ vs. pathloss-to-noise ratio $S$.}
    \label{fig:Q5vsS}
\end{figure}

Fig. \ref{fig:Q5vsS} shows the ultimate outage probability $\mathrm{Pr}_{\mathrm{out}}(M)$ vs. the pathloss-noise-ratio $S$ for the different power allocation methods. Since the infinite blocklength method overestimates achievable rate and hence underestimates the outage probability, it tends to allocate lower power than necessary given the outage probability. Observe from the figure that the infinite blocklength method fails to meet the outage probability constraint of $10^{-4}$. The max power allocation has the opposite behavior. It satisfies the outage probability constraint and yet squanders the power, with the resulting outage probability being way below the target. A desirable balance is attained by the proposed power allocation method. It controls the power allocation precisely so that the resulting outage probability is fairly close to the target $10^{-4}$.

\begin{figure}
    \centering    \centerline{\includegraphics[width=0.9\linewidth]{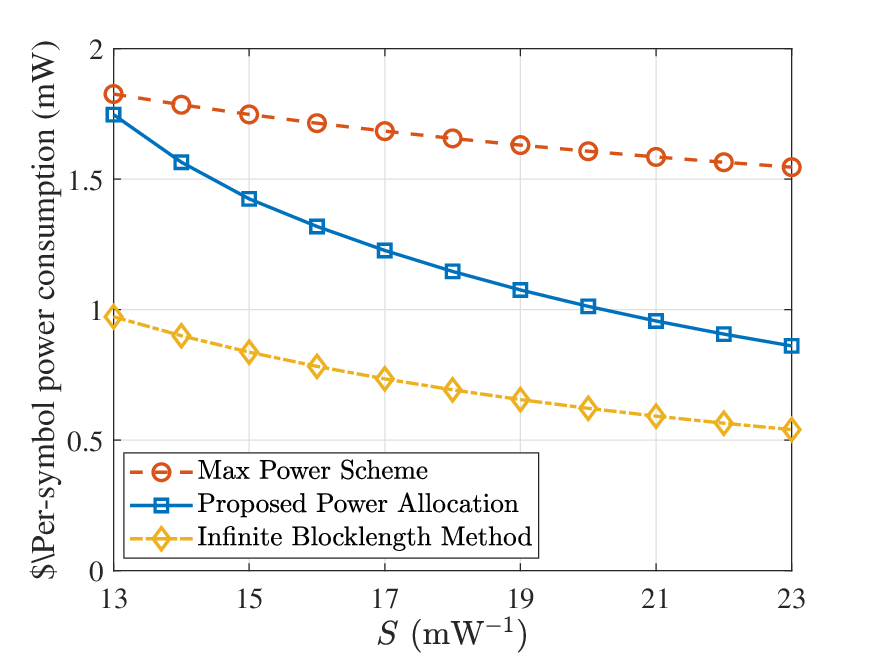}}
    \caption{Per-symbol power consumption $\frac{E}{L}$ vs. pathloss-to-noise ratio $S$.}
    \label{fig:EvsS}
\end{figure}

Fig. \ref{fig:EvsS} shows the per-symbol power consumption $E/L$ vs. the pathloss-to-noise ratio $S$ for the different methods. As compared with the max power scheme, the proposed method cuts down the energy consumption significantly, e.g., about $45\%$ energy is saved when $S = 23\text{ mW}^{-1}$. Observe that the infinite blocklength method requires slightly lower energy than the proposed method because of its underestimation of the outage probability. However, these two methods cannot guarantee the outage probability constraint as formerly shown in Fig. \ref{fig:Q5vsS}.

\section{Conclusion}
The aim of this work is to write the outage probability as an analytical function of the power variables in the finite-blocklength regime, and thereby optimize the power allocation for the IR-HARQ efficiently in those ultra-reliable
application scenarios. Toward this end, we propose a novel analytic upper bound on the outage probability, based on which the original power control problem can be converted to the GP problem. According to the simulation results, the power allocation optimized by the proposed GP-based algorithm requires much lower energy consumption while meeting the high reliability 
constraints.

\bibliographystyle{IEEEtran}
\bibliography{IEEEabrv,strings}                       

\begin{thebibliography}{10}
\providecommand{\url}[1]{#1}
\csname url@samestyle\endcsname
\providecommand{\newblock}{\relax}
\providecommand{\bibinfo}[2]{#2}
\providecommand{\BIBentrySTDinterwordspacing}{\spaceskip=0pt\relax}
\providecommand{\BIBentryALTinterwordstretchfactor}{4}
\providecommand{\BIBentryALTinterwordspacing}{\spaceskip=\fontdimen2\font plus
\BIBentryALTinterwordstretchfactor\fontdimen3\font minus
  \fontdimen4\font\relax}
\providecommand{\BIBforeignlanguage}[2]{{%
\expandafter\ifx\csname l@#1\endcsname\relax
\typeout{** WARNING: IEEEtran.bst: No hyphenation pattern has been}%
\typeout{** loaded for the language `#1'. Using the pattern for}%
\typeout{** the default language instead.}%
\else
\language=\csname l@#1\endcsname
\fi
#2}}
\providecommand{\BIBdecl}{\relax}
\BIBdecl

\bibitem{ldpcirharq}
I.~Andriyanova and E.~Soljanin, ``Optimized {IR-HARQ} schemes based on
  punctured {LDPC} codes over the {BEC},'' \emph{{IEEE} Trans. Inf. Theory},
  vol.~58, no.~10, pp. 6433--6445, Oct. 2012.

\bibitem{caire}
G.~Caire and D.~Tuninetti, ``The throughput of hybrid-{ARQ} protocols for the
  {G}aussian collision channel,'' \emph{{IEEE} Trans. Inf. Theory}, vol.~47,
  no.~5, pp. 1971--1988, Jul. 2001.

\bibitem{lanemanbound}
J.~Laneman, ``Limiting analysis of outage probabilities for diversity schemes
  in fading channels,'' in \emph{Proc. {IEEE} Global Commun. Conf. (GLOBECOM)},
  vol.~3, Dec. 2003.

\bibitem{shen}
K.~Shen, W.~Yu, X.~Chen, and S.~R. Khosravirad, ``Energy efficient {HARQ} for
  ultrareliability via novel outage probability bound and geometric
  programming,'' \emph{{IEEE} Trans. Wireless Commun.}, vol.~21, no.~9, pp.
  7810--7820, Apr. 2022.

\bibitem{wenyubound}
W.~Wang and K.~Shen, ``A new outage probability bound for {IR-HARQ} and its
  application to power adaptation,'' in \emph{2022 IEEE 23rd Int. Workshop
  Signal Process. Advances Wireless Commun. (SPAWC)}, Jul. 2022.

\bibitem{urllcfinite}
H.~Ren, C.~Pan, Y.~Deng, M.~Elkashlan, and A.~Nallanathan, ``Resource
  allocation for secure {URLLC} in mission-critical {IoT} scenarios,''
  \emph{{IEEE} Trans. Commun.}, vol.~68, no.~9, pp. 5793--5807, Sep. 2020.

\bibitem{buckingham}
D.~Buckingham and M.~C. Valenti, ``The information-outage probability of
  finite-length codes over {AWGN} channels,'' in \emph{2008 42nd Annual
  Conference on Information Sciences and Systems}, Mar. 2008.

\bibitem{polyanskiy}
Y.~Polyanskiy, H.~V. Poor, and S.~Verdu, ``Channel coding rate in the finite
  blocklength regime,'' \emph{{IEEE} Trans. Inf. Theory}, vol.~56, no.~5, pp.
  2307--2359, May 2010.

\bibitem{predictor}
H.~Guo, B.~Makki, and T.~Svensson, ``Predictor antennas for moving relays:
  Finite block-length analysis,'' in \emph{2020 3rd Int. Conf. Advanced Comm.
  Technol. Netw. (CommNet)}, 2020.

\bibitem{throughput1}
B.~Makki, T.~Svensson, and M.~Zorzi, ``Finite block-length analysis of the
  incremental redundancy {HARQ},'' \emph{{IEEE} Wireless Commun. Lett.},
  vol.~3, no.~5, pp. 529--532, Oct. 2014.

\bibitem{throughput2}
Y.~Li, M.~C. Gursoy, and S.~Velipasalar, ``Throughput of {HARQ-IR} with finite
  blocklength codes and {QoS} constraints,'' in \emph{Proc. {IEEE} Int. Symp.
  Inf. Theory (ISIT)}, Jun. 2017.

\bibitem{energy1}
M.~Centenaro, G.~Ministeri, and L.~Vangelista, ``A comparison of
  energy-efficient {HARQ} protocols for {M2M} communication in the finite
  block-length regime,'' in \emph{2015 {IEEE} Int. Conf. Ubiquitous Wireless
  Broadband (ICUWB)}, Oct. 2015.

\bibitem{he}
F.~He, Z.~Shi, B.~Zhou, G.~Yang, X.~Li, X.~Ye, and S.~Ma, ``{BLER} analysis and
  optimal power allocation of {HARQ-IR} for mission-critical {IoT}
  communications,'' \emph{{IEEE} Internet Things J.}, Aug. 2024.

\bibitem{avranas}
A.~Avranas, M.~Kountouris, and P.~Ciblat, ``Energy-latency tradeoff in
  ultra-reliable low-latency communication with retransmissions,'' \emph{{IEEE}
  J. Sel. Areas Commun.}, vol.~36, no.~11, pp. 2475--2485, 2018.

\bibitem{oCSIfading}
K.~N. Le, ``{HARQ} under outdated {CSI}, finite-blocklength regimes and
  generalised-{R}ician fading,'' \emph{IEEE Trans. Veh. Technol.}, vol.~71,
  no.~3, pp. 3298--3302, 2022.

\bibitem{diffpow_lambound1}
W.~Su, S.~Lee, D.~A. Pados, and J.~D. Matyjas, ``Optimal power assignment for
  minimizing the average total transmission power in hybrid-{ARQ} {R}ayleigh
  fading links,'' \emph{{IEEE} Trans. Commun.}, vol.~59, no.~7, pp. 1867--1877,
  Jul. 2011.

\bibitem{diffpow_lambound2}
T.~V.~K. Chaitanya and E.~G. Larsson, ``Outage-optimal power allocation for
  hybrid {ARQ} with incremental redundancy,'' \emph{{IEEE} Trans. Wireless
  Commun.}, vol.~10, no.~7, pp. 2069--2074, Jul. 2011.

\bibitem{laneman2006distribution}
J.~N. Laneman, ``On the distribution of mutual information,'' in \emph{Inf.
  Theory Appl. Workshop}, 2006.

\bibitem{DAHLMAN2018251}
E.~Dahlman, S.~Parkvall, and J.~Sköld, in \emph{5G NR: The Next Generation
  Wireless Access Technology}.\hskip 1em plus 0.5em minus 0.4em\relax Academic
  Press, 2018.

\end{thebibliography}

\end{document}